\documentclass[journal]{IEEEtran}

\usepackage{cite}

\ifCLASSINFOpdf
  \usepackage[pdftex]{graphicx}
\else
  \usepackage[dvips]{graphicx}
\fi

\usepackage[cmex10]{amsmath}

\usepackage{eufrak}

\usepackage{nameref}

\newcommand{\vect}[1]{\boldsymbol{\mathbf{#1}}}

\begin{document}

\title{An Analysis of How Spatiotemporal Dynamic Models of Brain Activity Could Improve MEG/EEG Inverse Solutions}

\author{Camilo~Lamus,
        Matti~S.~H{\"a}m{\"a}l{\"a}inen,
        Emery~N.~Brown,~\IEEEmembership{Fellow,~IEEE}
        and~Patrick~L.~Purdon,
\thanks{C. Lamus is with the Department of Brain and Cognitive Sciences, Massachusetts Institute of Technology, Cambridge, MA 02139 USA (e-mail: camilo@neurostat.mit.edu).}
\thanks{M.S. H{\"a}m{\"a}l{\"a}inen is with Harvard Medical School, Boston, MA 02115 USA, with the Athinoula A. Martinos Center for Biomedical Imaging, Department of Radiology, Massachusetts General Hospital, Charlestown, MA 02129 USA, and with the Department of Neuroscience and Biomedical Engineering, Aalto University School of Science, Espoo, Finland.}
\thanks{E.N. Brown is with the Department of Brain and Cognitive Sciences and the Institute for Medical Engineering and Science, Massachusetts Institute of Technology, Cambridge, MA 02139 USA, with the Department of Anesthesia, Critical Care and Pain Medicine, Massachusetts General Hospital, Boston, MA 02114 USA, and with Harvard Medical School, Boston, MA 02115 USA.}
\thanks{P.L. Purdon is with Harvard Medical School, Boston, MA 02115 USA, and with the Department of Anesthesia, Critical Care and Pain Medicine, Massachusetts General Hospital, Boston, MA 02114 USA (e-mail: patrickp@nmr.mgh.harvard.edu)}
\thanks{Manuscript received November 11, 2015; revised Month Day, Year.}}

\markboth{Preprint,~Vol.~x, No.~y, November~2015}%
{Lamus \MakeLowercase{\textit{et al.}}: Dynamic Lead Field Mapping}

\maketitle

\begin{abstract}

MEG and EEG are noninvasive functional neuroimaging techniques that provide recordings of brain activity with high temporal resolution, and thus provide a unique window to study fast time-scale neural dynamics in humans. However, the accuracy of brain activity estimates resulting from these data is limited mainly because 1) the number of sensors is much smaller than the number of sources, and 2) the low sensitivity of the recording device to deep or radially oriented sources. These factors limit the number of sources that can be recovered and bias estimates to superficial cortical areas, resulting in the need to include a priori information about the source activity. The question of how to specify this information and how it might lead to improved solutions remains a critical open problem. In this paper we show that the incorporation of knowledge about the brain’s underlying connectivity and spatiotemporal dynamics could dramatically improve inverse solutions. To do this, we develop the concept of the \textit{dynamic lead field mapping}, which expresses how information about source activity at a given time is mapped not only to the immediate measurement, but to a time series of measurements. With this mapping we show that the number of source parameters that can be recovered could increase by up to a factor of ${\sim20}$, and that such improvement is primarily represented by deep cortical areas. Our result implies that future developments in MEG/EEG analysis that model spatialtemporal dynamics have the potential to dramatically increase source resolution.

\end{abstract}


\IEEEpeerreviewmaketitle

\section{Introduction}
\label{sec:intro}

\IEEEPARstart{M}{agnetoencephalography} (MEG) and electroencephalography (EEG) are functional neuroimaging tools that provide noninvasive recordings of the magnetic and electric fields at the scalp generated by neuronal currents. MEG and EEG hold particular promise as tools to noninvasively study fast time-scale neural dynamics of human brain function in health and disease~\cite{Hamalainen:1993ws}. This makes MEG/EEG unique amongst other functional neuroimaging techniques such as functional magnetic resonance imaging (fMRI), positron emission tomography (PET), or diffuse optical tomography (DOT), which instead provide indirect measures of brain activity related to slower neurovascular changes. However, to properly interpret MEG and EEG recordings and fully realize their potential, one needs to estimate the source current activity underlying the measured electric potentials and magnetic fields at the scalp surface, i.e., to solve the ill-posed MEG/EEG inverse problem. Although significant progress has been made on this problem in the past few decades ~\cite{Hamalainen:1994uk,PascualMarqui:1994tp,Nummenmaa:2007gy,Wipf:2009gj,Baillet:1997ue,Friston:2008jr,Ou:2009cm,Bolstad:2009eg,Gramfort:2013en,Galka:2004ef,Long:2011fn,Lamus:2012gp}, at present MEG/EEG inverse solutions are are rough estimates with poor spatial resolution (in the order of a few cm), that are insensitive to the majority of deep cortical and subcortical regions.

The accuracy of source estimates is limited in part due to two main factors: the dimensionality and the biophysics in MEG/EEG inverse problem. These factors can be understood by analyzing the \textit{lead field matrix} ${\vect{X}}$, which maps the cortical activity of a few thousand dipole sources, ${\vect{\beta}_t}$, to the recordings in a few hundred scalp sensors, ${\vect{y}_t}$, at an individual time instant ${t}$~\cite{Hamalainen:1993ws,Dale:1993wo}:

\begin{equation*} \label{eq:leadfield}
    \vect{y}_t = \vect{X} \vect{\beta}_t + \textit{``noise''}.
\end{equation*}

\noindent In the case of the dimensionality, since the number of sources to be estimated is an order of magnitude larger than the number of sensors, different source configurations can produce identical scalp recordings, making solutions to this inverse problem non-unique. As a result, the number of sources that we can expect to recover from measurements in an individual time point ${t}$, which is determined by the rank of ${\vect{X}}$, is much smaller than the number of sources. The second factor is due to the biophysics of MEG/EEG. The amplitude of the scalp fields and potentials rapidly decays with the inverse square of the distance from the sensor to the source~\cite{Hamalainen:1993ws}. Therefore, the sensitivity of the recording device---the signal power generated by an individual active source across all sensors given by the norm of the columns of ${\vect{X}}$---is very low for a large percentage of cortical areas. The geometry of the electromagnetic fields generated by current dipoles poses additional challenges. The measured fields tend to be widely distributed across the scalp, and radially-oriented sources can be magnetically silent~\cite{Ahlfors:2010bs}. Overall, the dimensionality and biophysics underlying MEG and EEG limit the number of source parameters that can be recovered, and effectively restrict estimates to cortical areas whose activity is most easily detected by the sensors, ultimately imposing the need to include a priori information about the source currents. The question of how to specify this a priori information and how it might improve solutions is one of the most fundamental problems in bioelectromagnetism and neuroimaging.

Prior information about the current dipole distribution has been used in the MEG/EEG inverse problem literature in order to obtain unique estimates. This information has taken the form of a probabilistic model or optimization penalty that implicitly or explicitly assumes cortical activity is either independent across time~\cite{Hamalainen:1994uk,PascualMarqui:1994tp,Nummenmaa:2007gy,Wipf:2009gj}, temporally or spatio-temporally smooth~\cite{Baillet:1997ue,Friston:2008jr,Ou:2009cm,Bolstad:2009eg,Gramfort:2013en}, or follow a linear dynamic process~\cite{Galka:2004ef,Long:2011fn,Lamus:2012gp}. While these priors alleviate issues related to the non-uniqueness of source estimates, they do not necessarily improve on the limitations that stem from the rank deficiency and restricted sensitivity of the lead field matrix. However, one way to ameliorate these issues could be achieved by incorporating prior knowledge about the spatiotemporal physiological relations present in brain activity~\cite{Bullock:1995wy,Destexhe:1999tc,Leopold:2003va,Nunez:1995vc}. This is because in such a spatiotemporal system, information about source activation at a given time instant ${t}$ is contained not only in the immediate measurement ${\vect{y}_t}$ but also in multiple observations over a time interval ${[\vect{y}_1,\vect{y}_2,\ldots,\vect{y}_T]}$. As a result of this dynamic flow of information, we can imagine a mapping that relates brain activity at a given moment in time to the measurements in the complete analysis interval---a \textit{dynamic} lead field mapping. The rank and sensitivity of the dynamic lead field mapping could be substantially better than those of the static lead field matrix ${\vect{X}}$, not only because of its increased dimensionality, but also because of how information can flow from brain areas that are harder to detect at the scalp to those that are easier to detect.

Inspired by electrophysiology and neuroanatomy studies~\cite{Bullock:1995wy,Destexhe:1999tc,Leopold:2003va,Nunez:1995vc}, in this paper we show how the incorporation of information about the source spatial connectivity and dynamics could dramatically increase both the number of sources that can be recovered, as well as the sensitivity for detecting such sources. To do this, we develop the concept of the \textit{dynamic lead field mapping}, a dynamic extension of the lead field matrix that allows us to analyze the rank and sensitivity of the mapping between source activity at a given time instant to the complete set of measurements both forward and backwards in time. This dynamic lead field mapping, though developed for the specific problem of MEG/EEG source imaging, is firmly grounded in dynamic systems theory. With this mapping we show that the number of sources that can be effectively recovered increases by up to a factor of ${\sim 20}$ by modeling the most basic local cortical dynamic connections. Furthermore, we show that the inclusion of such local cortical connections increases the sensitivity for detecting sources distributed across the brain, and that the increase in sensitivity is more pronounced in sources located within sulci and other deeper areas. At the core of our technical development is a projection operation that allows us to analyze the rank and sensitivity resulting from different prior source models. In particular, we analyze a space-time separable model and a static model that assumes temporal independence, both of which are frequently used in MEG/EEG analysis. We find that improvements in the number and sensitivity of sources that can be recovered occur only when dynamic spatiotemporal connections are modeled. Our results imply that future developments in MEG/EEG analysis that explicitly model dynamic connections between brain areas have the potential to dramatically increase spatiotemporal resolution by taking full advantage of the brain's inherent spatiotemporal structure, connectivity, and dynamics.

\section{Methods}
\label{sec:methods}

\subsection{The spatiotemporal dynamic source model and the MEG/EEG measurement model}
\label{subsec:dynmodel}

Converging lines of evidence suggest that brain activity is a spatiotemporal dynamic process that is organized in part by structural connections at different scales. At the smaller spatial scale, intracranial recordings in different species have revealed that cortical activity exhibits strong correlations that persist up to a distance of ${10 \: \mathrm{mm}}$ during rest and task periods~\cite{Bullock:1995wy,Destexhe:1999tc,Leopold:2003va}. These local cortico-cortical dynamic interactions can be supported neuroanatomically by the fact that axonal collateral projections from pyramidal cells spread laterally approximately ${6 \:\mathrm{mm}}$ along the cortical surface~\cite{Nunez:1995vc}. In addition, long-distance correlations can exist by means of white matter tracts that connect distant brain regions, like those which are thought to support large-scale brain network activity such as the resting-state networks studied via fMRI~\cite{Hagmann:2008gd,Honey:2009eh,vandenHeuvel:2009gc}.

In order to provide a conservative analysis where spatiotemporal dynamics consistent with neurophysiological evidence are considered, we choose to focus on a parsimonious model that incorporates local spatiotemporal interactions. One way to model local spatiotemporal connections of this type is to use a first order linear dynamic process. In this model, neuronal currents at a given point in time ${t}$ and spatial location ${i}$, ${\beta_{i,t}}$, are a function of past neuronal currents at the same location, ${\beta_{i,t-1}}$, as well as past currents, ${\beta_{j,t-1}}$, at locations ${j}$ within a local neighborhood ${\mathcal{N}(i)}$:

\begin{equation} \label{eq:sourcedyn}
    \beta_{i,t} = f_{i,i} \beta_{i,t-1} + \sum_{j \in \mathcal{N}(i)} f_{i,j} \beta_{j,t-1} + \omega_{i,t}.
\end{equation}

\noindent In Equation~\eqref{eq:sourcedyn}, the weights ${f_{i,j}}$ represent the interaction between sources at locations $i$ and $j$. At each location $i$, the weights $f_{i,j}$ corresponding to its neighbors ${j\in \mathcal{N}(i)}$ are assumed to be positive and inversely proportional to the distance between sources. In addition, they are normalized such that the contribution of the neighbors to the dynamics of the ${i}$th source equals its self contribution, while the total contribution is constant $0 < \phi < 1$ in order to obtain stable source dynamics, i.e., ${\sum_{j\in \mathcal{N}(i)}f_{i,j} = f_{i,i}}$ and ${\sum_{j\in \mathcal{N}(i)}f_{i,j} + f_{i,i} = \phi < 1}$. Furthermore, the input process ${\omega_{i,t}}$ is assumed to be Gaussian with zero-mean and independent across time and space. The spatiotemporal model in Equation~\eqref{eq:sourcedyn}, which has been previously used in dynamic source localization analysis~\cite{Galka:2004ef,Lamus:2012gp}, can be readily expressed in vector form as:

\begin{equation} \label{eq:state}
    \vect{\beta}_t = \vect{F}\vect{\beta}_{t-1} + \vect{\omega}_t,
\end{equation}

\noindent where ${\vect{\beta}_t}$ is a vector of dimension ${p}$ (${\sim \: 10^3}$), and the input process ${\vect{\omega}_t}$ is Gaussian with zero mean and independent across time with spatial covariance matrix $\vect{Q}$, which we assume to be diagonal and positive definite.

In an MEG/EEG experiment, we obtain a recording of the magnetic field and electric potential from hundreds of sensors located on or above the scalp at times ${t\in[1,2,\ldots ,T]}$. At time ${t}$, the resulting data vector ${\vect{y}_t}$, which is of dimension ${n}$ (${\sim \: 10^2}$), is related to the source vector ${\vect{\beta}_t}$ by the observation equation~\cite{Hamalainen:1993ws,Dale:1993wo}:

\begin{equation} \label{eq:obsmodel}
    \vect{y}_t = \underbrace{\vect{X}\vect{\beta}_t}_{\text{Signal}} + \underbrace{\vect{\varepsilon}_t}_{\text{Noise}},
\end{equation}

\noindent where ${\vect{X}}$ is the ${n \times p}$ lead field matrix computed using a quasistatic approximation of the Maxwell's equations, and ${\vect{\varepsilon}_t}$ is the Gaussian white noise vector with zero mean and spatial covariance $\vect{R}$ representing background instrumental and environmental noise. In Equation~\eqref{eq:obsmodel}, ${\vect{X}\vect{\beta}_t}$ represents the \textit{signal} portion of the model, and the \textit{noise} term ${\vect{\varepsilon}_t}$ is independent from ${\vect{\beta}_t}$ for all time points.

\subsection{The dynamic lead field mapping}
\label{subsec:dynleadfield}

\begin{figure}[htb]
\begin{center}
\includegraphics[width=\columnwidth]{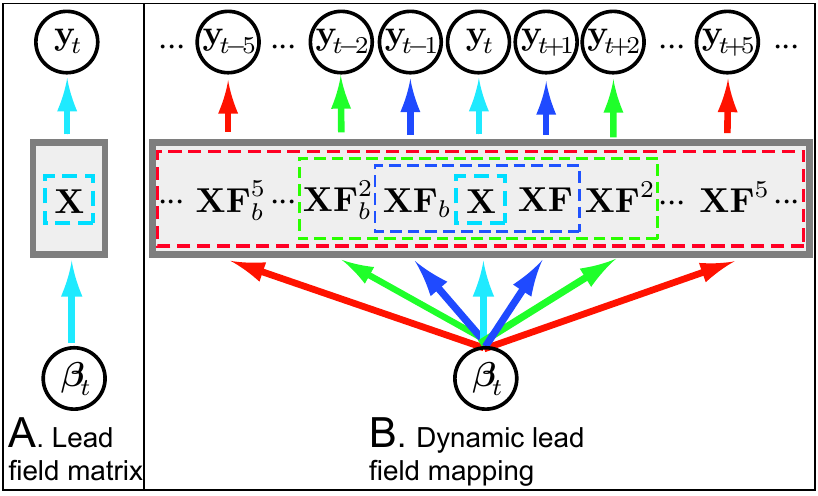}
\end{center}
\caption{{\bf The dynamic lead field mapping.}
A: The (static) lead field matrix $\vect{X}$ determines how information from the source vector $\vect{\beta}_t$ propagates to the measurement $\vect{y}_t$ at the same point in time $t$. B: When spatiotemporal dynamics are modeled, the dynamic lead field mapping $\vect{D}_t$ determines how information from the source vector $\vect{\beta}_t$ at a given time $t$ is mapped to the complete time series of measurements $[\vect{y}_1,\vect{y}_2, \ldots, \vect{y}_T]$.}
\label{fig:dynleadfield}
\end{figure}

In the observation model (Eq.~\ref{eq:obsmodel}), the signals generated by brain activity are represented by the product of the lead field matrix and the source vector: ${\vect{X}\vect{\beta}_t}$. Since the resulting product and the source vector are independent of the noise term ${\vect{\varepsilon}_t}$, the lead field matrix ${\vect{X}}$ contains all the information related to the mapping of the brain source vector ${\vect{\beta}_t}$ at a particular point in time ${t}$ to the measurement ${\vect{y}_t}$ at that same point in time ${t}$ (Figure~\ref{fig:dynleadfield}A). From this static point of view, it is clear that the maximum number of independent variables that could be determined from a single measurement in time is limited by the rank of the ${\vect{X}}$ matrix, which unfortunately is less than or equal to the number of sensors, and much smaller than the number of sources~\cite{Foucart:2013wp}. However, if we consider the brain's source activity as a spatiotemporal dynamic process, we can dramatically improve on these limitations.

In a spatiotemporal system, the information about the source vector ${\vect{\beta}_t}$ at a particular time ${t}$ is contained not only in the immediate observation ${\vect{y}_t}$ but also in the previous ${[\vect{y}_1,\vect{y}_2,\ldots,\vect{y}_{t-1}]}$ and future ${[\vect{y}_{t+1},\vect{y}_{t+2},\ldots,\vect{y}_T]}$ observations. Because of this, information from the source vector ${\vect{\beta}_t}$ is effectively mapped to the complete time series of measurements through a function (Figure~\ref{fig:dynleadfield}B), which we call the \textit{dynamic lead field mapping} ${\vect{D}_t}$, whose rank and sensitivity would be greater than those of the static lead field matrix ${\vect{X}}$. As we will show below, this mapping can be expressed as:

\begin{equation} \label{eq:dynleadfield}
    \begin{bmatrix}
        \vect{y}_1 \\
        \vect{y}_2 \\
        \vdots \\
        \vect{y}_T
    \end{bmatrix}
    = \underbrace{\vect{D}_t\vect{\beta}_t}_{\text{Signal}} + \underbrace{\vect{n}_t}_{\text{Noise}},
\end{equation}

\noindent where the noise term ${\vect{n}_t}$ has dimension ${nT}$. In analogy to the lead field matrix ${\vect{X}}$ in the \textit{static} observation model (Eq.~\ref{eq:obsmodel}), the \textit{dynamic} lead field mapping $\vect{D}_t$ can be used to determine the number of independent variables we can recover as well as the sensitivity for detecting such sources when we consider the complete time series of observations in a spatiotemporal dynamic framework.

To obtain the dynamic lead field mapping ${\vect{D}_t}$ in Equation~\eqref{eq:dynleadfield}, we must derive an observation model for the complete set of measurements ${[\vect{y}_1,\vect{y}_2,\ldots,\vect{y}_T]}$ where: i) the \textit{signal} portion of the model is a function of only the source vector ${\vect{\beta}_t}$ at a particular time ${t}$, and ii) the \textit{noise} term is independent of the source vector ${\vect{\beta}_t}$. We will consider three separate cases depending on whether the observation vector ${\vect{y}_{t\pm k}}$ ($k = 0, 1, \ldots$) corresponds to a future, present, or past observation with respect to the source vector ${\vect{\beta}_t}$ at the present time ${t}$. Because of this partition we will consider the dynamic lead field mapping ${\vect{D}_t}$ as a block matrix composed of submatrices ${\vect{P}_{t\pm k,t}}$:

\begin{equation*} \label{eq:dynleadproj}
    \vect{D}_t = \begin{bmatrix}
                        \vect{P}_{1,t} \\
                        \vect{P}_{2,t} \\
                        \vdots \\
                        \vect{P}_{T,t}
                    \end{bmatrix}.
\end{equation*}

\noindent We call the blocks ${\vect{P}_{t\pm k,t}}$ the projection matrices.

In the case where the measurement vector corresponds to the present observation (${k=0}$), we use the static observation model (Eq.~\ref{eq:obsmodel}) to obtain the relation between ${\vect{\beta}_t}$ and ${\vect{y}_t}$ in the block number ${t}$ of Equation~\eqref{eq:dynleadfield}:

\begin{equation*} \label{eq:dynleadpresentblock}
    \vect{y}_t = \underbrace{\vect{X}\vect{\beta}_t}_{\text{Signal}} + \underbrace{\vect{\varepsilon}_t}_{\text{Noise}}.
\end{equation*}

\noindent From the equation above we can identify that ${\vect{P}_{t,t}=\vect{X}}$.

To obtain the projection matrices corresponding to future observations, we start by using the recursion defining the source vector spatiotemporal dynamics (Eq.~\ref{eq:state}) and iterate it ${k\in[1,2,\ldots]}$ times to obtain:

{\small
\begin{align} \label{eq:kfuturepred}
    \vect{\beta}_{t+1} &= \vect{F} \vect{\beta}_t + \vect{\omega}_{t+1}  \nonumber\\
    \vect{\beta}_{t+2} &= \vect{F}^2 \vect{\beta}_t + \vect{F}\vect{\omega}_{t+1} + \vect{\omega}_{t+2} \nonumber\\
    & \vdots \nonumber\\
    \vect{\beta}_{t+k} &= \vect{F}^k \vect{\beta}_t + \sum_{j=1}^k \vect{F}^{k-j}\vect{\omega}_{t+j}.
\end{align}
}%

\noindent Using this ${k}$-step iteration into the future along with the static measurement model (Eq.~\ref{eq:obsmodel}), we obtain the relation between ${\vect{\beta}_t}$ and the future measurement ${\vect{y}_{t+k}}$:

\begin{equation} \label{eq:dynleadfutureblock}
    \vect{y}_{t+k} = \underbrace{\vect{X}\vect{F}^k \vect{\beta}_t}_{\text{Signal}} + \underbrace{\sum_{j=1}^k \vect{X}\vect{F}^{k-j}\vect{\omega}_{t+j} + \vect{\varepsilon}_{t+k}}_{\text{Noise}}.
\end{equation}

\noindent We should note that Equation~\eqref{eq:dynleadfutureblock} represents the ${(t+k)}$th block of Equation~\eqref{eq:dynleadfield}, and therefore we identify that ${\vect{P}_{t+k,t}=\vect{X}\vect{F}^k}$.

To derive the past projection matrices we make use of an equivalent representation of the source dynamics (Eq.~\ref{eq:state}) known as the \textit{backwards Markovian} model~\cite{Kailath:2000vg}:

\begin{equation} \label{eq:backwardsmarkov}
    \vect{\beta}_t = \vect{F}_b \vect{\beta}_{t+1} + \vect{\omega}_{t}^{b}.
\end{equation}

\noindent In this representation, where the state vector ${\vect{\beta}_t}$ evolves backwards in time, the time-reversed transition matrix is given by ${\vect{F}_b = \vect{C}\vect{F}'\vect{C}^{-1}}$, where ${\vect{C}=\mathrm{Cov}(\vect{\beta}_t)}$ is the steady-state source covariance (see Section \ref{subsec:steadystate} in \nameref{sec:supplement} for details). The backwards input process ${\vect{\omega}_t^b}$, which is independent across time and Gaussian, has zero mean and covariance matrix ${\vect{Q}_b = (\vect{C} - \vect{F}_b\vect{C}\vect{F}_b')}$. With this equivalent backwards dynamic representation (Eq.~\ref{eq:backwardsmarkov}), we proceed to obtain the relation between ${\vect{\beta}_t}$ and the previous measurement ${\vect{y}_{t-k}}$ just as we did when we considered future measurement. In this case the ${k \in [1,2,\dots]}$ step backwards iteration is given by:

{\small
\begin{align} \label{eq:kpastpred}
    \vect{\beta}_{t-1} =& \vect{F}_b \vect{\beta}_t + \vect{\omega}_{t-1}^{b} \nonumber \\
    \vect{\beta}_{t-2} =& \vect{F}_b^2 \vect{\beta}_t + \vect{F}_b\vect{\omega}_{t-1}^{b} + \vect{\omega}_{t-2}^{b} \nonumber \\
    &\vdots \nonumber \\
    \vect{\beta}_{t-k} =& \vect{F}_b^k \vect{\beta}_t + \sum_{j=1}^k \vect{F}_b^{k-j}\vect{\omega}_{t-j}^b.
\end{align}
}%

\noindent We use the static observation model (Eq.~\ref{eq:obsmodel}) along with the ${k}$-step past iteration to obtain the ${(t-k)}$th block of Equation~\eqref{eq:dynleadfield}:

\begin{equation} \label{eq:dynleadpastblock}
    \vect{y}_{t-k} = \underbrace{\vect{X}\vect{F}_b^k \vect{\beta}_t}_{\text{Signal}} + \underbrace{\sum_{j=1}^k \vect{X}\vect{F}_b^{k-j}\vect{\omega}_{t-j}^b + \vect{\varepsilon}_{t-k}}_{\text{Noise}},
\end{equation}

\noindent where we see that ${\vect{P}_{t-k,t}=\vect{X}\vect{F}_b^k}$.

At this point we are ready to explicitly write the complete observation model (Eq.~\ref{eq:dynleadfield}) as well as the dynamic lead field mapping ${\vect{D}_t}$. For a particular time ${t}$, we use the equations mapping the source vector ${\vect{\beta}_t}$ to the present (Eq.~\ref{eq:obsmodel}), future (Eq.~\ref{eq:dynleadfutureblock}), and past (Eq.~\ref{eq:dynleadpastblock}) to obtain the complete observation model:

{\small
\begin{equation} \label{eq:dynleadfieldexplicit}
    \begin{bmatrix}
        \vect{y}_1 \\
        \vdots \\
        \vect{y}_{t-1} \\
        \vect{y}_t \\
        \vect{y}_{t+1} \\
        \vdots \\
        \vect{y}_T
    \end{bmatrix}
    =
    \underbrace{
    \begin{bmatrix}
        \vect{X}\vect{F}_b^{t-1} \\
        \vdots \\
        \vect{X}\vect{F}_b \\
        \vect{X} \\
        \vect{X}\vect{F} \\
        \vdots \\
        \vect{X}\vect{F}^{T-t} \\
    \end{bmatrix}}_{\vect{D}_t} \vect{\beta}_t
    +
    \underbrace{
    \begin{bmatrix}
        \vect{e}_{1,t} \\
        \vdots \\
        \vect{e}_{t-1,t} \\
        \vect{e}_{t,t} \\
        \vect{e}_{t+1,t} \\
        \vdots \\
        \vect{e}_{T,t}
    \end{bmatrix}}_{\vect{n}_t},
\end{equation}
}

\noindent where error terms are given by,

{\small
\begin{equation*} \label{eq:projerror}
    \vect{e}_{t \pm k,t} =
    \begin{cases}
        \sum_{j=1}^k  \vect{X}\vect{F}_b^{k-j}\vect{\omega}_{t-j}^b + \vect{\varepsilon}_{t-k},
        & -k>0 \\
        \vect{\varepsilon}_{t},
        & k = 0 \\
        \sum_{j=1}^k \vect{X}\vect{F}^{k-j}\vect{\omega}_{t+j} + \vect{\varepsilon}_{t+k},
        & +k>0
    \end{cases}.
\end{equation*}
}

From the complete observation model (Eq.~\ref{eq:dynleadfieldexplicit}) we can see that the dynamic lead field mapping is:

\begin{equation} \label{eq:dynleadfieldmapping}
    \vect{D}_t = \left[
        \vect{F}_b^{t-1'}\vect{X}', \hdots,
        \vect{F}_b'\vect{X}',
        \vect{X}',
        \vect{F}'\vect{X}', \hdots,
        \vect{F}^{T-t'}\vect{X}'\right]'.
\end{equation}

\noindent In addition, we can observe that Equation~\eqref{eq:dynleadfieldexplicit} expresses the complete set of measurements ${[\vect{y}_1,\vect{y}_2,\ldots,\vect{y}_T]}$ as the sum of the \textit{signal} function ${\vect{D}_t\vect{\beta}_t}$ and \textit{noise} term  ${\vect{n}_t = [\vect{e}_{1,t}', \vect{e}_{2,t}',\ldots,\vect{e}_{T,t}']'}$. We should note that the noise term ${\vect{n}_t}$ is indeed independent of the source vector ${\vect{\beta}_t}$, since the vectors ${\vect{e}_{t\pm k,t}}$ correspond to the projection errors that result from projecting the measurements ${\vect{y}_{t\pm k}}$ onto the source vector ${\vect{\beta}_t}$. Such projections are obtained via the matrices ${\vect{P}_{t\pm k,t}}$ (For details see Section~\ref{subsec:dynleadisproj} in~\nameref{sec:supplement}). Consequently, the \textit{dynamic} lead field mapping ${\vect{D}_t}$ contains all the information describing how the brain source vector ${\vect{\beta}_t}$, at a particular point in time ${t}$, propagates to the complete set of measurements. Therefore, we can use ${\vect{D}_t}$ to determine the number of independent variables we can recover, as well as the sensitivity for detecting such sources, from the complete time series of observations in a spatiotemporal dynamic framework.

\subsection{Extension to general Gaussian source models}
\label{subsec:equivdynleadfield}

The expression for the dynamic lead field $\vect{D}_t$ in Equation~\eqref{eq:dynleadfieldmapping} stems from a specific model of spatiotemporal dynamics. In this section we derive a more general expression for $\vect{D}_t$ for Gaussian spatiotemporal models. This will allow us to compare the properties of our dynamic source model to those of more commonly used models. In these more general models, the joint distribution of the sources has a zero mean and is Gaussian, as in our spatiotemporal dynamic source model (Eqs.~\ref{eq:sourcedyn} and~\ref{eq:state}), but the spatiotemporal covariance structure differs from that in our model. Specifically, under a model ${\mathfrak{m}}$, the joint distribution of the source vectors ${[\vect{\beta}_1^{(\mathfrak{m})}, \vect{\beta}_2^{(\mathfrak{m})}, \ldots, \vect{\beta}_T^{(\mathfrak{m})}]}$ is Gaussian with zero mean (${\mathrm{E}[\vect{\beta}_t^{(\mathfrak{m})}]=\vect{0}}$) and arbitrary cross-covariances ${\mathrm{E}[\vect{\beta}_{k}^{(\mathfrak{m})}\vect{\beta}_t^{(\mathfrak{m})'}]}$, where $\mathrm{E}$ denotes the expectation operator. The MEG/EEG measurement are obtained via the static observation model (Eq.~\ref{eq:obsmodel}):

\begin{equation} \label{eq:equivobsmodel}
    \vect{y}_t^{(\mathfrak{m})} = \vect{X}\vect{\beta}_t^{(\mathfrak{m})} + \vect{\varepsilon}_t.
\end{equation}

Just as we did in the case of the dynamic lead field mapping, when considering a model ${\mathfrak{m}}$ we would like to express the complete time series of measurements ${[\vect{y}_1^{(\mathfrak{m})}, \vect{y}_2^{(\mathfrak{m})}, \ldots, \vect{y}_T^{(\mathfrak{m})}]}$ as the sum of a function that only depends on the source vector ${\vect{\beta}_t^{(\mathfrak{m})}}$ at the present time ${t}$ and a noise term that is independent of ${\vect{\beta}_t^{(\mathfrak{m})}}$. As we will show below, this can be done via the linear relation,

\begin{equation} \label{eq:equivdynleadfield}
    \begin{bmatrix}
        \vect{y}_1^{(\mathfrak{m})} \\
        \vect{y}_2^{(\mathfrak{m})} \\
        \vdots \\
        \vect{y}_T^{(\mathfrak{m})}
    \end{bmatrix}
    = \underbrace{\vect{D}_t^{(\mathfrak{m})}\vect{\beta}_t^{(\mathfrak{m})}}_{\text{Signal}} + \underbrace{\vect{n}_t^{(\mathfrak{m})}}_{\text{Noise}},
\end{equation}

\noindent where ${\vect{D}_t^{(\mathfrak{m})}}$ is the equivalent of the dynamic lead field mapping for the model ${\mathfrak{m}}$, and the noise term ${\vect{n}_t^{(\mathfrak{m})}}$ is independent of the source vector ${\vect{\beta}_t^{(\mathfrak{m})}}$.

To obtain Equation~\eqref{eq:equivdynleadfield} we again consider the matrix ${\vect{D}_t^{(\mathfrak{m})}}$ as composed of sub-matrix blocks ${\vect{P}_{t\pm k,t}^{(\mathfrak{m})}}$:

\begin{equation*} \label{eq:equivdynleadproj}
    \vect{D}_t^{(m)} = \begin{bmatrix}
                        \vect{P}_{1,t}^{(\mathfrak{m})} \\
                        \vect{P}_{2,t}^{(\mathfrak{m})} \\
                        \vdots \\
                        \vect{P}_{T,t}^{(\mathfrak{m})}
                    \end{bmatrix}.
\end{equation*}

\noindent To obtain the ${t\pm k}$th block of Equation~\eqref{eq:equivdynleadfield} we make use of the fact that, under a model $\mathfrak{m}$, the measurement vector ${\vect{y}_{t\pm k}^{(\mathfrak{m})}}$ at time $t\pm k$ can be expressed as the sum of its projection onto the source vector ${\vect{\beta}_t^{(\mathfrak{m})}}$ at time ${t}$ and the projection error:

\begin{equation*} \label{eq:equivprojection}
    \vect{y}_{t\pm k}^{(\mathfrak{m})} = \vect{P}_{t\pm k,t}^{(\mathfrak{m})}\vect{\beta}_t^{(\mathfrak{m})} + \vect{e}_{t\pm k,t}^{(\mathfrak{m})},
\end{equation*}

\noindent where the projection error ${\vect{e}_{t\pm k,t}^{(\mathfrak{m})}}$ is independent of the source vector at time ${t}$, and the projection matrix ${\vect{P}_{t\pm k,t}^{(\mathfrak{m})}}$ is the solution of the orthogonality equation\footnote{The projection error ${\vect{e}_{t\pm k,t}^{(\mathfrak{m})}=\vect{y}_{t\pm k}^{(\mathfrak{m})}-\vect{P}_{t\pm k,t}^{(\mathfrak{m})}\vect{\beta}_t^{(\mathfrak{m})}}$ is uncorrelated with ${\vect{\beta}_t^{(\mathfrak{m})}}$, namely ${\mathrm{E}[(\vect{y}_{t\pm k}^{(\mathfrak{m})}-\vect{P}_{t\pm k,t}^{(\mathfrak{m})}\vect{\beta}_t^{(\mathfrak{m})})\vect{\beta}_t^{(\mathfrak{m})'}]=\vect{0}}$~\cite{Kailath:2000vg}. Since ${\vect{y}_{t\pm k}^{(\mathfrak{m})}}$ and ${\vect{\beta}_t^{(\mathfrak{m})}}$ are jointly Gaussian, the projection error ${\vect{e}_{t\pm k,t}^{(\mathfrak{m})}}$ and ${\vect{\beta}_t^{(\mathfrak{m})}}$ are also jointly Gaussian. Therefore, the projection error and the source vector are uncorrelated and jointly Gaussian, and thus are independent.}:

\begin{equation} \label{eq:equivorthogonalitycond}
    \mathrm{E}\left[(\vect{y}_{t\pm k}^{(\mathfrak{m})}-\vect{P}_{t\pm k,t}^{(\mathfrak{m})}\vect{\beta}_t^{(\mathfrak{m})})\vect{\beta}_t^{(\mathfrak{m})'}\right] = \vect{0}.
\end{equation}

\noindent By making use of the static observation model (Eq.~\ref{eq:equivobsmodel}), it is easy to see that the projection matrices that solve the orthogonality equation (Eq.~\ref{eq:equivorthogonalitycond}) under a given source model ${\mathfrak{m}}$ are given by:

\begin{equation} \label{eq:equivprojmatrix}
    \vect{P}_{t\pm k,t}^{(\mathfrak{m})} = \vect{X}\left(\mathrm{E}[\vect{\beta}_{t\pm k}^{(\mathfrak{m})}\vect{\beta}_{t}^{(\mathfrak{m})'}]\right)\left(\mathrm{E}[\vect{\beta}_{t}^{(\mathfrak{m})}\vect{\beta}_{t}^{(\mathfrak{m})'}]\right)^{-1}.
\end{equation}

\noindent Therefore, for a given Gaussian model ${\mathfrak{m}}$ with a specific spatiotemporal source covariance structure, we can compute the corresponding projection matrices ${\vect{P}_{t\pm k,t}^{(\mathfrak{m})}}$ and, just as we did in the development of the dynamic lead field mapping (Eqs.~\ref{eq:dynleadfield} and~\ref{eq:dynleadfieldexplicit}), obtain an equivalent observation model for the complete series of measurements by stacking up the projection matrices.

To give concrete examples of the ${\vect{D}_t^{(\mathfrak{m})}}$ matrix for general Gaussian source models, we will analyze two models commonly used in the source localization literature: i) the first, which we denote as the \textit{IND} model, assumes the source vectors are independent across time~\cite{Hamalainen:1994uk}, i.e., ${\mathrm{E}[\vect{\beta}_{t\pm k}^{(ind)}\vect{\beta}_t^{(ind)'}]=\vect{0}}$ for ${k \neq 0}$; and ii) a space-time separable (\textit{STS}) model in which the joint source covariance factors via the Kronecker product ${\vect{\Gamma} \otimes \vect{C}}$ into a purely spatial covariance matrix ${\vect{C}}$ and a purely temporal covariance matrix ${\vect{\Gamma}}$~\cite{Friston:2008jr}. We should note that in the \textit{STS} model, the covariance between source vectors at times ${t\pm k}$ and ${t}$ is given by ${\mathrm{E}[\vect{\beta}_{t\pm k}^{(sts)}\vect{\beta}_t^{(sts)'}]=\gamma_{t\pm k,t}\vect{C}}$, where ${\gamma_{k,t}}$ is the element of the temporal covariance matrix ${\vect{\Gamma}}$ at position ${t\pm k,t}$. With the covariance structure defined for these two models, and using Equation~\eqref{eq:equivprojmatrix}, we can see that the projection matrices for the \textit{IND} and \textit{STS} models are

\begin{equation*} \label{eq:equivprojectionmatrixexplicit}
    \vect{P}_{t\pm k,t}^{(ind)} =
        \begin{cases}
            \vect{X} \quad k = 0 \\
            \vect{0} \quad k \neq 0
        \end{cases}
    \quad
    \text{and}
    \quad
    \vect{P}_{t\pm k,t}^{(sts)} = \frac{\gamma_{t\pm k,t}}{\gamma_{t,t}}\vect{X},
\end{equation*}

\noindent respectively. Therefore, the ${\vect{D}_t^{(\mathfrak{m})}}$ matrix for the \textit{IND} and \textit{STS} models, which constitute the equivalent of the ${\vect{D}_t}$ matrix for these Gaussian models, are given by:

\begin{equation} \label{eq:indstscompleteobsmodel}
    \vect{D}_t^{(ind)} =  \vect{1}_t \otimes \vect{X}
    \quad \text{and} \quad
    \vect{D}_t^{(sts)} =  \frac{1}{\gamma_{t,t}}\vect{\gamma}_t \otimes \vect{X}
\end{equation}

\noindent respectively, where ${\vect{1}_t}$ is the unit vector with the $t$th entry set equal to one, and $\vect{\gamma}_t$ is the $t$th column of $\vect{\Gamma}$

\section{Results}
\label{sec:results}

\subsection{Rank and singular value spectrum of the dynamic lead field mapping ${\vect{D}_t}$}
\label{subsec:ranksingvaldynleadfield}

We used the dynamic lead field mapping construct in conjunction with an MRI-based MEG forward model from a human subject (see Section \ref{subsec:ethics} and \ref{subsec:datapreprocessing} in \nameref{sec:supplement}) to estimate the number of independent sources we can recover in a model that includes local spatiotemporal cortical dynamics (Eq.~\ref{eq:sourcedyn}). In order to avoid the computational challenges associated with finding the rank and singular values of the very large ${\vect{D}_t}$ matrix (Eq.~\ref{eq:dynleadfieldmapping}), we chose to analyze the truncated versions of ${\vect{D}_t}$ that correspond to a model that includes ${k \in [1,2,5,10,20]}$ measurements into the past and future of a given time ${t}$, ${[\vect{y}_{t-k},\ldots,\vect{y}_t,\ldots,\vect{y}_{t+k}]}$. The resulting mappings, which we denote by ${\vect{D}_{t}(k)}$, are given by:

\begin{equation} \label{eq:dynleadfieldtrunk}
    \vect{D}_{t}(k) = \left[
        \vect{F}_b^{k'}\vect{X}', \hdots,
        \vect{F}_b'\vect{X}', \vect{X}', \vect{F}'\vect{X}', \hdots
        \vect{F}^{k'}\vect{X}'\right]'.
\end{equation}

\noindent This arrangement was chosen to compare the relative contributions of increasing numbers of observations to the number of independent sources that can be effectively recovered. Furthermore, since the matrices ${\vect{D}_{t}(k)}$ correspond to a truncated version of the much larger matrix ${\vect{D}_t}$, our computations serve as a lower bound on the number of sources that could be recovered from the time series of measurements in the complete interval ${[1,2,\ldots,T]}$.

\begin{figure}[htb]
\begin{center}
\includegraphics[width=\columnwidth]{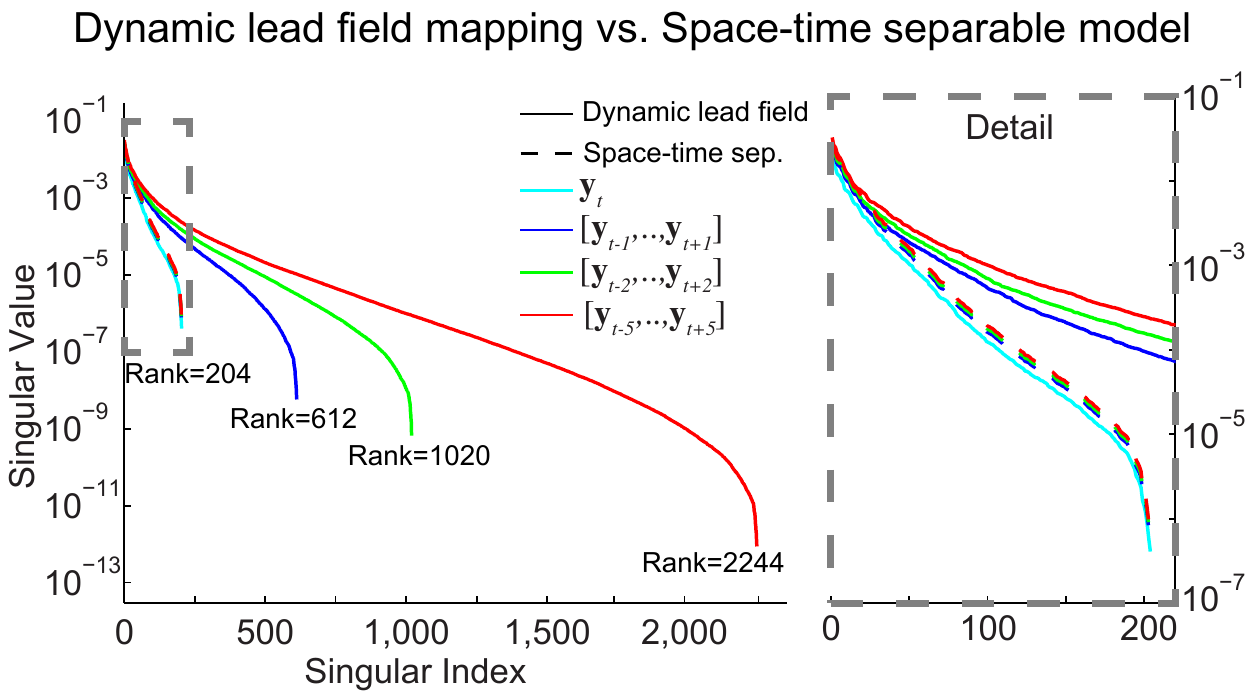}
\end{center}
\caption{{\bf Spectrum of the dynamic lead field mapping $\vect{D}_t(k)$.}
The evolution of the singular value spectra of $\vect{D}_t(k)$ (solid lines) with increasing number of temporal measurements $k$. The spectrum of this matrix increases thus indicating an increase in the number of sources that could be recovered in the spatiotemporal dynamic model with local connections. A zoom-in is shown (right) to detail the singular value spectra of ${\vect{D}_t(k)^{(sts)}}$ for the space-time separable (\textit{STS}) source model (dashed line). In the \textit{STS} source model the number of sources that could be recovered does not increase since the rank stays constant.}
\label{fig:dynleadspectum}
\end{figure}

Figure~\ref{fig:dynleadspectum} (left) and Table~\ref{tab:rankdynlead} show the evolution of the singular value spectrum and the rank of the ${\vect{D}_{t}(k)}$ matrix, respectively, as the number of observed data points ${2k+1}$ increases. With each successive increase in ${k}$, the singular value spectrum and the rank of the ${\vect{D}_{t}(k)}$ increases monotonically, reaching a value of ${\mathrm{rank}(\vect{D}_{t}(20))=4551}$. This indicates the number of sources that could be recovered increases by up to a factor of $20$ by modeling only local cortical dynamic connections.

\begin{table}[!ht]
\caption{
{\bf Rank of $\vect{D}_t(k)$.}}
\begin{tabular}{|c||c|c|c|c|c|c|}
\hline
$k$                               &   0   &   1   &   2   &   5       &   10      &   20      \\ \hline
$\mathrm{rank}(\vect{D}_t(k))$  &   204 &   612 &   1020&   2244    &   3889    &   4551    \\ \hline
\end{tabular}
\label{tab:rankdynlead}
\end{table}

Similarly, we analyzed the the number of sources that could be recovered in the \textit{IND} and \textit{STS} models by evaluating the rank and singular values spectrum of the ${\vect{D}_t^{(ind)}}$ and ${\vect{D}_t^{(sts)}}$ matrices, respectively (Eq.~\ref{eq:indstscompleteobsmodel}). Just as we did in the case of our spatiotemporal dynamic model, we evaluated the truncated versions of these matrices given by:

{\small
\begin{align} \label{eq:equivcompleteobsmodeltrunk}
    \vect{D}_t(k)^{(ind)} &= \left[
        \vect{0}', \hdots, \vect{0}', \vect{X}', \vect{0}', \hdots, \vect{0}'\right]' \\
    \vect{D}_t(k)^{(sts)} &= \nonumber \\
        &\hspace*{-0.5in}\frac{1}{\gamma_{t,t}}\left[
        \gamma_{t-k,t}\vect{X}', \hdots,
        \gamma_{t-1,t}\vect{X}',
        \vect{X}',
        \gamma_{t+1,t}\vect{X}', \hdots
        \gamma_{t+k,t}\vect{X}'\right] \nonumber
\end{align}
}%

\noindent For both the \textit{IND} and \textit{STS} models, the rank of these matrices did not increase with the inclusion of more temporal measurements: ${\mathrm{rank}(\vect{D}_{t}(k)^{(ind)})=\mathrm{rank}(\vect{D}_{t}(k)^{(sts)})=204}$ for all values of ${k}$. In the case of the \textit{STS} model, we saw that the 204 singular values of ${\vect{D}_{t}(k)^{(sts)}}$ were slightly increased (Fig.~\ref{fig:dynleadspectum}, right). However, such increase was uniform across singular values, as it can be shown that the vector with the ordered singular values of ${\vect{D}_{t}(k)^{(sts)}}$ is equal to a scaled version of the vector with the singular values of ${\vect{X}}$: ${\mathrm{svd}(\vect{D}_{t}(k)^{(sts)}) \propto \mathrm{svd}(\vect{X})}$. Importantly, the fact that the rank of ${\vect{D}_{t}(k)^{(ind)}}$ and ${\vect{D}_{t}(k)^{(sts)}}$ does not increase can be seen from the Equation~\eqref{eq:equivcompleteobsmodeltrunk}. In both cases, these matrices are obtained by stacking the lead field matrix ${\vect{X}}$ above and below with either a matrix of zeros or scaled versions of ${\vect{X}}$.

\subsection{Sensitivity analysis of dynamic lead field mapping ${\vect{D}_t}$}
\label{subsec:sensitivitydynleadfield}

\begin{figure}[htb]
\begin{center}
\includegraphics[width=\columnwidth]{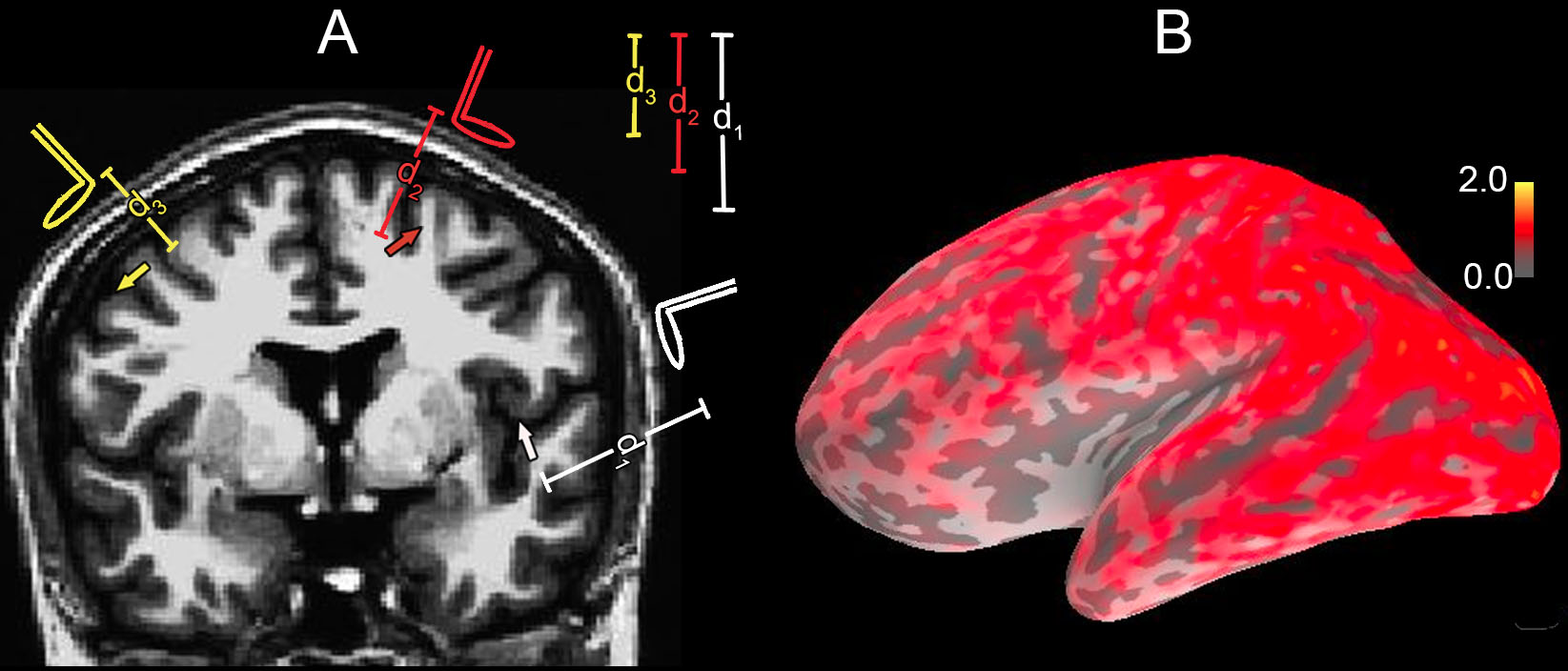}
\end{center}
\caption{{\bf Illustration of sensitivity of the static lead field matrix to dipoles at different depths.}
The white dipole located very deep in the right auditory cortex at a distance $d_1$ from its closest sensor produces very weak signals which makes it very difficult to detectable. The red dipole near the trough of a gyrus at a distance $d_2$ from its closest sensor produces relatively weak signals and is difficult to detect. The superficial yellow dipole in the side of a sulcus at a distance $d_3$ from its closest sensor produces strong signals. For each dipole, the absolute sensitivity can be computed using Equation~\eqref{eq:sensitivity} resulting in the sensitivity map overlaid on the inflated cortical surface (right). The dark and light gray areas on the inflated surface represent sulci and gyri, respectively. The color-scale is multiplied by $10^{-3}$.}
\label{fig:sensitivity}
\end{figure}

As discussed earlier, the biophysics of the MEG/EEG forward problem dictate that some regions of the brain are more difficult to observe and measure than others: the measured signal decays with the inverse of the square distance from the sensor to the source, and in the case of MEG, sources oriented near the radial direction are magnetically silent. Figure~\ref{fig:sensitivity} (left) illustrates this phenomenon where we show arrows representing dipole sources located in the right temporal cortex (deep dipole, white), the trough of a gyrus (mildly deep dipole, red), and the side of a gyrus (superficial dipole, yellow). In this hypothetical example, the deep white dipole located at a distance $d_1$ from its closest sensor produces very low amplitude signals which renders it very difficult to detect. The mildly deep red dipole at a distance $d_2$ from its closest sensor generates relatively weak signals, and this also makes it difficult to measure. In contrast to the previous two cases, the yellow dipole located on the superficial side of the gyrus at a distance $d_3$ from its closest sensor produces strong signals which makes it easier to recover. A natural way to quantify this ease or difficulty for detecting the electric and magnetic fields at the scalp surface when only the instantaneous ${\vect{y}_t}$ is observed, is to compute the signal power measured across sensors generated by a single active dipole source of unit amplitude. Specifically, if we fix the source vector to represent a unit amplitude active dipole at the $i$th cortical location, the total signal power it produces at the sensors is given by:

\begin{equation} \label{eq:sensitivity}
    s_{i,t}(0) = ||\vect{X}\vect{1}_i||_2,
\end{equation}

\noindent Performing such computation for all cortical locations ${i}$ results in a static profile of the \textit{absolute} sensitivity gain of the lead field matrix ${\vect{X}}$. Figure~\ref{fig:sensitivity} (right) shows this static sensitivity profile computed using lead field matrix described in the previous section. The sensitivity is overlaid on the inflated cortical surface where dark and light gray areas represent sulci and gyri, respectively. We saw that the sensitivity was highest in some portions of gyri and the more superficial cortical areas (red tones), but it was low in the troughs of sulci, insula, and inferior-frontal regions.

While the sensitivity of the lead field matrix gives us a static image of the signal quality obtained in the instantaneous measurement $\vect{y}_t$, it does not inform us of the quality of the measured signals in time when we account for the spatiotemporal dynamics of the underlying cortical activity. One way to asses such signal quality is to extend the sensitivity analysis done for the lead field matrix (Eq.~\ref{eq:sensitivity}) to the case of the dynamic lead field ${\vect{D}_t(k)}$. Specifically, if we assume a unit amplitude dipole is active at time ${t}$ and cortical location ${i}$ in our spatio-temporal dynamic model (Eq.~\ref{eq:state}), the total measured power in the time interval ${[t-k,\ldots,t+k]}$ across all sensor is given by:

\begin{equation} \label{eq:dynsensitivity}
     s_{i,t}(k) = ||\vect{D}_{t}(k)\vect{1}_i||_2,
\end{equation}

\begin{figure}[htb]
\begin{center}
\includegraphics[width=0.84\columnwidth]{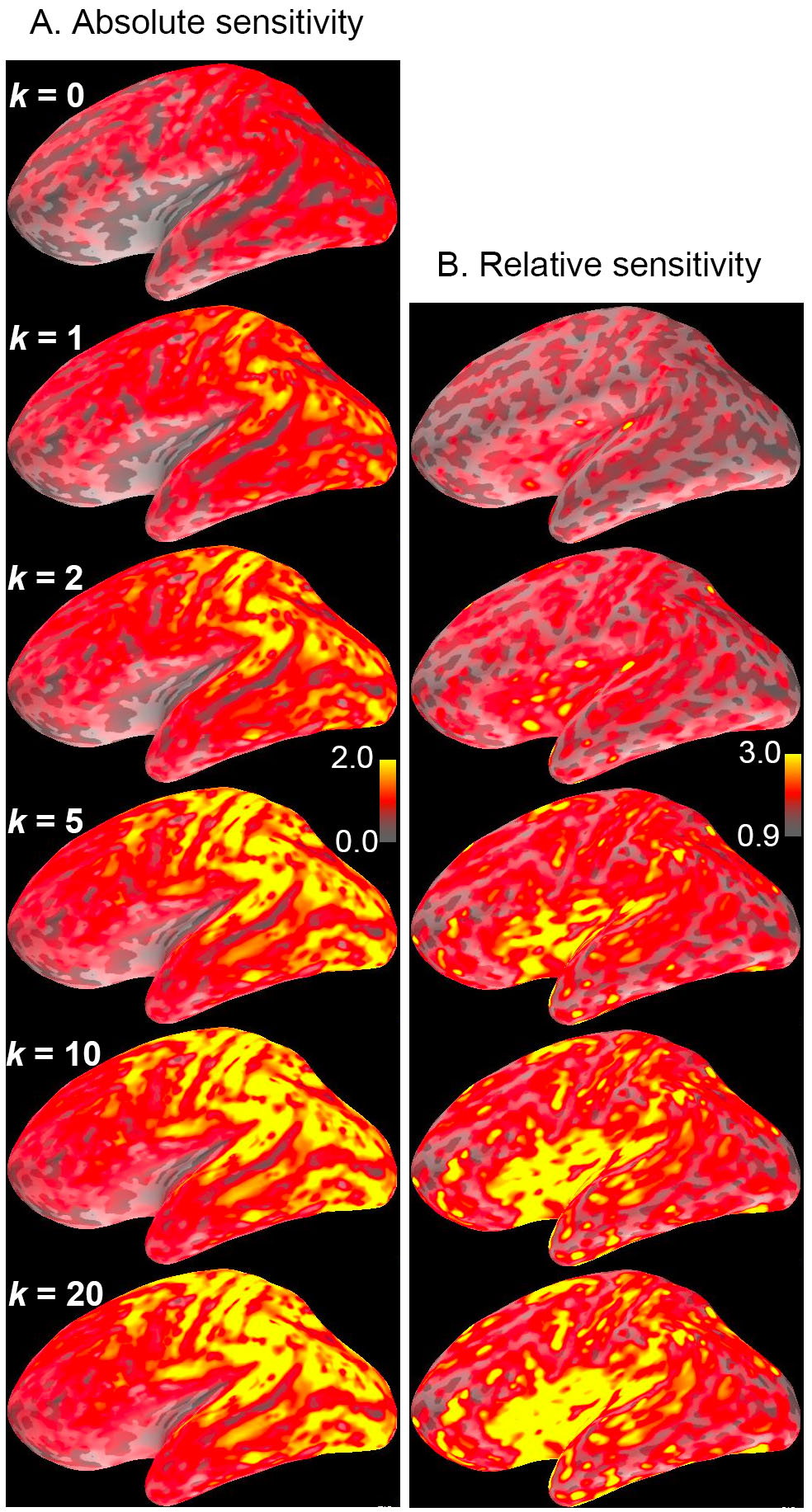}
\end{center}
\caption{{\bf Sensitivity analysis of the dynamic lead field mapping.}
A: Absolute sensitivity of ${\vect{D}_t(k)}$ as a function of the number of incorporated measurements ${2k + 1}$. When ${k=0}$, the sensitivity of the static lead field matrix ${\vect{X}=\vect{D}_t(0)}$ is shown. For ${k \geq 1}$, the sensitivity increases in most cortical regions including sulci. The color-scale is multiplied by $10^{-3}$. B: The relative sensitivity of the dynamic lead field mapping increases with $k$ in most cortical regions. This relative increase is more pronounced in areas with low static sensitivity such as sulci, insula, and inferior frontal cortex.}
\label{fig:dynsensitivity}
\end{figure}

\noindent Figure~\ref{fig:dynsensitivity}A (left panels) shows the absolute sensitivity of the dynamic lead field mapping as we increase the number of measurement in time ($2k+1$). We saw that the sensitivity increased in most cortical regions including sulci when more temporal information is included, but this increase was not as pronounced in deeper regions such as some portions of insula and the inferior-frontal cortex.

To quantify the improvement in the dynamic lead field sensitivity ${s_{i,t}(k)}$ in relation to the static case, i.e., the static lead field matrix sensitivity ${s_{i,t}(0)}$, we computed the \textit{relative} sensitivity gain: ${s_{i,t}(k)/s_{i,t}(0)}$. Figure~\ref{fig:dynsensitivity}B (right panels) shows the relative sensitivity gain of the dynamic lead field mapping. We found that the relative sensitivity increases in the majority of cortical regions with additional temporal information. Interestingly, the increments in relative sensitivity were higher in regions of low absolute sensitivity, such as sulci, insula, and inferior frontal cortex (yellow tones).

\begin{figure}[htb]
\begin{center}
\includegraphics[width=0.84\columnwidth]{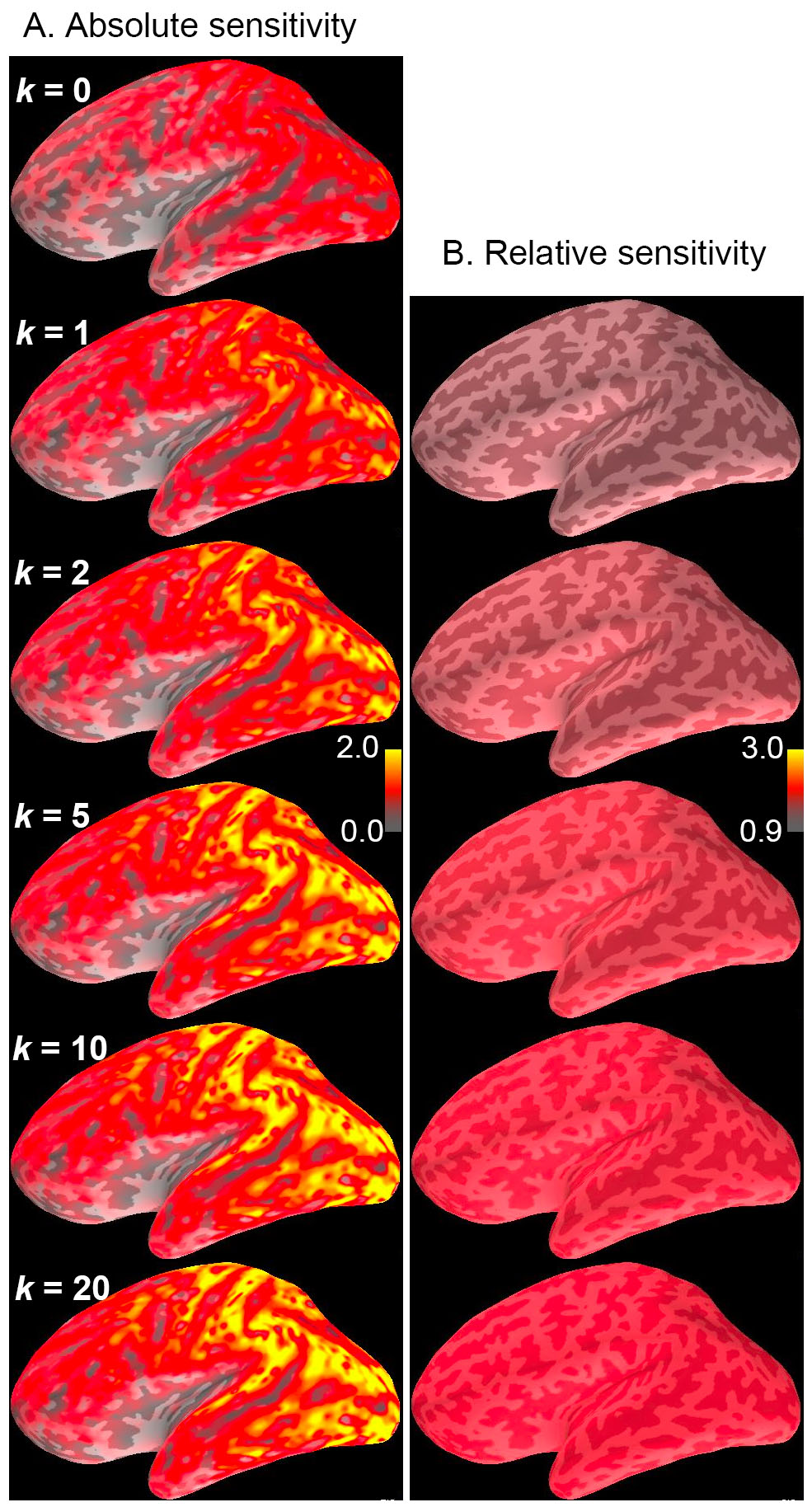}
\end{center}
\caption{{\bf Sensitivity analysis of the space-time separable model.}
A: Absolute sensitivity for the space-time separable model ${\vect{D}_t(k)^{(sts)}}$ as a function of $k$. The sensitivity increases in this case but is not as widespread nor as high as it was in the case of the dynamic lead field (Fig.~\ref{fig:dynsensitivity}A). The color-scale is multiplied by $10^{-3}$. B: The increments in relative sensitivity for the space-time separable model are spatially uniform and not as pronounced in regions that are difficult to detect such as sulci, insula, and inferior frontal cortex.}
\label{fig:stssensitivity}
\end{figure}

We performed an equivalent sensitivity analysis assuming that the cortical activation was generated by the space-time separable (\textit{STS}) model. To compute this sensitivity, we used Equation~\eqref{eq:dynsensitivity} but in this case replaced ${\vect{D}_{t}(k)}$ with the truncated matrix ${\vect{D}_t(k)^{(sts)}}$ (Eq.~\ref{eq:equivcompleteobsmodeltrunk}). Figure~\ref{fig:stssensitivity}A (left panels) shows the absolute sensitivity obtained in the \textit{STS} model. We saw an increase in sensitivity in most cortical regions, but this increase did not appear as broadly spread as it was in the case of the dynamic lead field mapping (Fig.~\ref{fig:dynsensitivity}A). Figure~\ref{fig:stssensitivity}B shows the relative sensitivity in the (\textit{STS}) model. The relative sensitivity in the \textit{STS} model increased with the number of temporal measurements, but this increase was spatially uniform across cortex. Compared to the relative sensitivity of the dynamic lead field (Fig.~\ref{fig:dynsensitivity}B), the increments in relative sensitivity under the \textit{STS} model were not as pronounced in regions that are more difficult to detect from a static point of view, such as sulci, insula, and inferior-frontal regions.

\begin{figure}[htb]
\begin{center}
\includegraphics[width=0.42\columnwidth]{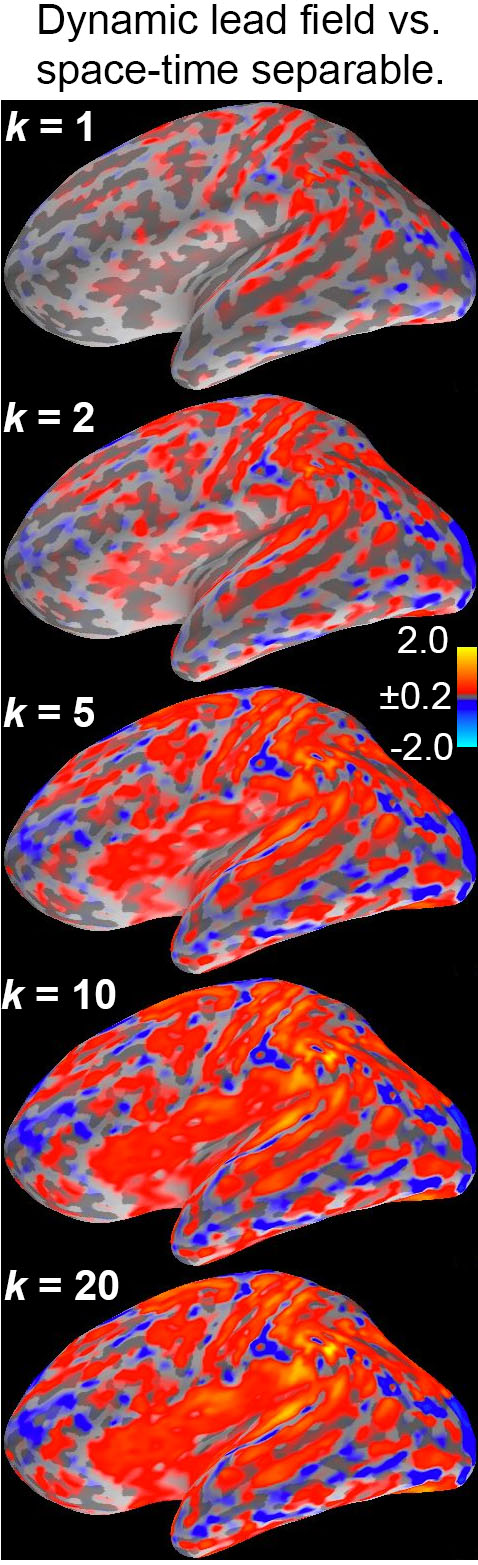}
\end{center}
\caption{{\bf Difference in absolute sensitivity between dynamic lead field and space-time separable models.}
For the majority of cortical areas the sensitivity in the dynamic lead field mapping is higher than that of the \textit{STS} model (red and yellow tones). For ${k \geq 5}$, the difference in sensitivity is more pronounced in favor of the dynamic lead field mapping in deeper cortical areas. The color-scale is multiplied by $10^{-3}$.}
\label{fig:diffsensitivity}
\end{figure}

Figure~\ref{fig:diffsensitivity} shows the difference in absolute sensitivity between our spatiotemporal dynamic model and the \textit{STS} model for each value of $k$: ${s_{i,t}(k) - s_{i,t}(k)^{(sts)}}$. We should note that the color scale in this case is different from that in panel A of Figures~\ref{fig:dynsensitivity} and~\ref{fig:stssensitivity}: the red and blue tones indicate small positive and negative differences of ${\sim 0.2\times10^{-3}}$, respectively; large positive difference of ${\sim 2\times10^{-3}}$ are shown in yellow, while large negative differences are shown in light blue. For the majority of cortical areas, and all values of $k$, the sensitivity was higher in the spatiotemporal dynamic model as indicated by the red and yellow tones. The small areas where the sensitivity of the \textit{STS} model was slightly higher (blue areas where the difference is ${\sim -0.2\times10^{-3}}$) were mainly represented by superficial portions of sulci which are indeed already detectable from a static point of view, i.e., if we only consider the immediate measurement for analysis. Interestingly, for values of ${k \geq 5}$ the sensitivity in the spatiotemporal dynamic model is higher in deeper areas such as sulci and insular cortex, which are precisely the areas that are more difficult to detect from a static point of view.

\begin{figure}[htb]
\centering
\includegraphics[width=0.84\columnwidth]{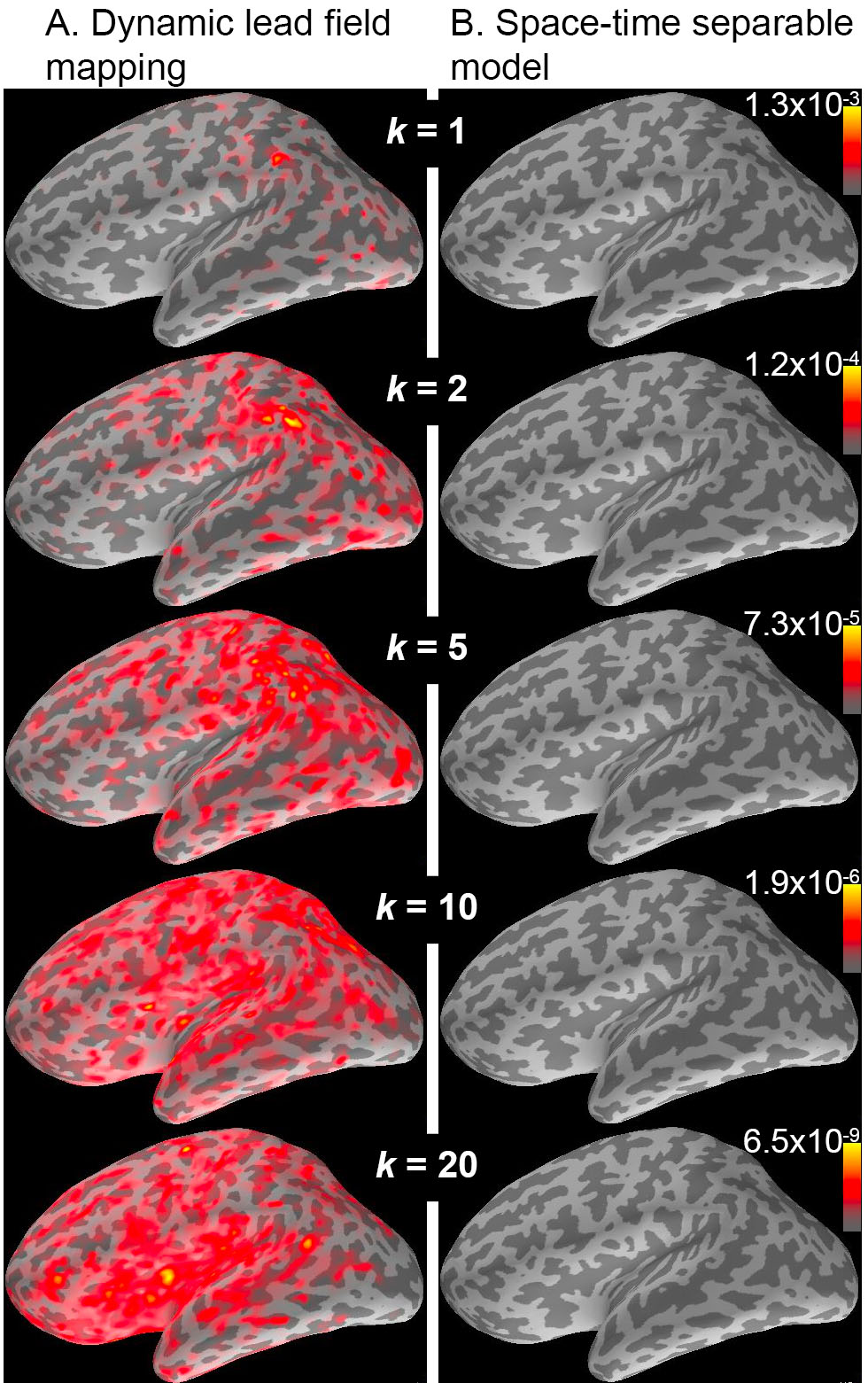}
\caption{{\bf Sensivity in newly accessible dimensions: Dynamic lead field vs. space-time separable model.}
A: The sensitivity of the dynamic lead field mapping in the newly accessible dimension increases as a function of $k$ and is highest in some portions of sulci, insula, and inferior frontal regions. B: In the case of the space-time separable model, there are no sensitivity gains in previously unaccessible dimensions since the rank of the matrix ${\vect{D}_t(k)^{(sts)}}$ does not increase.}
\label{fig:perpsensitivity}
\end{figure}

We saw in the previous section that, in a model that includes local spatiotemporal cortical dynamics, adding temporal measurements (${k}$) increases the rank of the ${\vect{D}_{t}(k)}$ matrices. We therefore wanted to characterize the dynamic lead field sensitivity that corresponded to the rank increments, which thus represent the sensitivity gains in the newly accessible dimensions. Specifically, if we consider two dynamic lead field matrices ${\vect{D}_{t}(k_1)}$ and ${\vect{D}_{t}(k_2)}$ with ${k_1<k_2}$, the sensitivity in the dimension accessible at ${k_2}$ but not at ${k_1}$ can be obtained by: first projecting the rows of ${\vect{D}_{t}(k_2)}$ onto the null space of ${\vect{D}_{t}(k_1)}$; and then compute the sensitivity of the matrix constructed with the projected rows. Figure~\ref{fig:perpsensitivity}A shows the sensitivity in the newly accessible areas for the pairs of values corresponding to consecutive number in ${k \in [1,2,5,10,20]}$. As $k$ increased, the distribution of sensitivity for the newly accessible dimensions was highest in some portions of sulci, insula, and inferior-frontal regions, which are poorly detected areas in the static case (Fig.~\ref{fig:perpsensitivity}A). We performed an equivalent analysis for the \textit{STS} model. In this case there were no gains in sensitivity in the previously unaccessible dimensions (Fig.~\ref{fig:perpsensitivity}B). This results from the fact that the rank of the ${\vect{D}_{t}(k)^{(sts)}}$ matrix does not increase irrespective to the number of analyzed measurements ($k$).

\section{Conclusions and discussion}
\label{sec:discussion}

In this work we have shown that the number of independent sources that can be recovered from MEG recordings can increase up to a factor of ${\sim20}$ by modeling linear source dynamics and local cortical connections. To do this, we developed and analyzed the concept of the \textit{dynamic lead field mapping}. This dynamic mapping extends the static observation model that originates from the lead field biophysics (Eq.~\ref{eq:obsmodel}) to to account for spatiotemporal dynamics~\cite{Bullock:1995wy,Destexhe:1999tc,Leopold:2003va,Nunez:1995vc,Hagmann:2008gd,Honey:2009eh,vandenHeuvel:2009gc}. The dynamic lead field mapping expresses the relation between the cortical source vector at any given time $t$ and the MEG/EEG measurements in the complete experiment interval ${[\vect{y}_1, \vect{y}_2, \ldots, \vect{y}_T]}$, creating an observation model for the complete time series of measurements that maintains the crucial signal-plus-noise structure present in its static counterpart. Since the rank of the lead field matrix determines the number of independent variables we can expect to recover from an individual measurement in time, the rank of the dynamic lead field mapping determines the number of variables we can recover in the dynamic case from the complete time series of observations. Therefore, in a typical MRI-based source model with $\sim 5124$ dipole sources and $\sim 204$ sensors, while the static (\textit{IND}) or space-time separable (\textit{STS}) models can recover at most ${\sim 4\% \;(204/5124\times 100)}$ of the independent variables, the approximate ${4551/204 \approx 20}$ fold rank improvement of the dynamic lead field implies that we could potentially recover up to ${\sim 89\% \;(4551/5124\times 100)}$ of the independent variables if we were to model the brain's source activity as a spatiotemporal dynamic process.

While the static lead field matrix provides information about the source vector from a single measurement at a given point in time, the dynamic lead field mapping provides additional information from past and future measurements that reflect the trajectory of the dynamic cortical state. Along this trajectory, activity hidden within a ``blind spot'' of the static lead field at a given time can evolve into a ``visible'' portion of the static lead field due to the brain's high level of connectivity. Since the dynamic lead field mapping explicitly describes how the measurement at one time point contains information about the cortical state at another time point, this mapping effectively shows how information can flow from blind spots in the static lead field to more visible locations.  In principle, depending on the cortical connectivity and dynamics, it is then possible to access and potentially estimate a much larger number of independent variables. Our work shows that modeling the simplest spatiotemporal relationships could dramatically improve source localization.  Moreover, it suggests that more comprehensive models of connectivity and spatiotemporal dynamics could provide even greater improvements.

We performed a sensitivity analysis of the dynamic lead field mapping to determine how sensitivity in different cortical areas changes as we increased the number of measurements in time. This analysis showed that, in our spatiotemporal dynamic paradigm, brain areas that are difficult to detect from a static point of view, such as those with low sensitivity in troughs of sulci, insula, and inferior-frontal regions---``blind spots'' as described above---become more accessible with the addition of temporal measurements. Moreover, we showed that this increase in sensitivity in deep sources does not occur in static (\textit{IND}) or space-time separable (\textit{STS}) models, suggesting that activity within ``blind spots'' become accessible only through their connection to more visible regions. This could explain the improvements in MEG/EEG source localization accuracy obtained via algorithms that explicitly model source dynamics with local cortical connections~\cite{Galka:2004ef,Lamus:2012gp}.

We envision a possible way in which our dynamic lead field construct could be extended to more general source models, to potentially include non-Gaussian models or more realistic spatiotemporal neural source dynamics. In principle, this could be achieved by taking techniques from control theory designed to analyze deterministic systems, and adapting them to characterize properties of stochastic dynamical systems. In control theory, a basic question is to determine if different initial states of a deterministic dynamical system necessarily produce different measurements. When this is the case, the system is called \textit{observable}, since the system's initial states can be distinguished based on measurements alone~\cite{Kailath:2000vg}. In our linear dynamic state-space model (Eqs.~\ref{eq:state} and~\ref{eq:obsmodel}), for example, assuming that the state input ${\vect{\omega}_t}$ is a priori known and deterministic, and that the measurements are noiseless (${\vect{\varepsilon}_t=0}$), effectively converts our model into a deterministic system. In this deterministic case, an unknown initial state ${\vect{\beta}_0}$ of the system can be distinguished based on noiseless measurements when the rank of the so-called observability matrix $\mathcal{O}={[\vect{X}', \vect{F}'\vect{X}', \ldots, \vect{F}^{p-1'}\vect{X}']'}$ is equal to the dimension of the state (${p}$). We can see here that the observability matrix $\mathcal{O}$ is related to the mapping given by the dynamic lead field from the state ${\vect{\beta}_t}$ to only the present and future observations ${[\vect{y}_t,\vect{y}_{t+1}, \ldots, \vect{y}_{t+p-1}]}$ (this can be seen in Equation~\eqref{eq:dynleadfieldmapping}). We therefore hypothesize that the dynamic lead field concept developed here could be extended by incorporating insights from the analysis of observability in deterministic nonlinear systems~\cite{Sontag:1998up}, and by applying the time-reversed representation of nonlinear stochastic dynamic systems~\cite{Anderson:1982hm}.

A long-standing conjecture in functional neuroimaging is that the integration of different imaging modalities, such as MEG, EEG, fMRI, and DOT, could improve spatiotemporal resolution, due to the complementary nature of the physics and physiology underlying these modalities. Extensions of our analysis to multimodal neuroimaging could provide a formal paradigm to characterize and maximize the spatiotemporal resolution that can be obtained from multimodal data. For example, an immediate multimodal effort could be aimed at including long-distance connections---both between distant cortical locations, and between cortex and low sensitivity subcortical areas---by using anatomical connectivity derived from diffusion MRI~\cite{Hagmann:2008gd}. We predict that the inclusion of these long-distance connections in a stochastic dynamics framework could have a dramatic impact on the number and location of source parameters that could be recovered from MEG/EEG time series data. While the focus of our work is in the field of neuroimaging, our dynamic lead field approach could have application in other areas where an underlying spatiotemporal stochastic dynamical system is observed through noisy serial measurements. Therefore, we hypothesize that our approach could find applications in areas such as geophysics, network theory, and epidemiology, where physical spatiotemporal dynamic relationships may exist, but are not yet exploited in inverse solutions.

Our analysis and results have been focused on the number, spatial distribution, and sensitivity of sources that can be recovered under a spatiotemporal dynamic framework. In practical applications, model misspecification, as well as the signal-to-noise of the measurements, will likely limit performance gains to some extent below the levels we have reported. However, the large improvements in rank and sensitivity that we have observed, even under a very simple spatiotemporal model, suggest that substantial performance improvements would be achievable in practical source localization applications. In future work, we will analyze source localization performance under this spatiotemporal framework and characterize the influence of different spatiotemporal model choices.


%



\section*{Acknowledgment}

We would like to thank Dr. Simona Temereanca for providing the MEG and MRI data used in this paper.

\ifCLASSOPTIONcaptionsoff
  \newpage
\fi



\bibliographystyle{IEEEtran}
\bibliography{papers}
\newpage
\onecolumn

\setcounter{equation}{0}
\setcounter{figure}{0}
\setcounter{table}{0}
\setcounter{page}{1}
\makeatletter
\renewcommand{\theequation}{S\arabic{equation}}
\renewcommand{\thefigure}{S\arabic{figure}}

\section{\bf Supplemental Materials}
\label{sec:supplement}

\begin{center}
{\bf An Analysis of How Spatiotemporal Dynamic Models of Brain Activity Could Improve MEG/EEG Inverse Solutions\\}
Camilo~Lamus, Matti~S.~H{\"a}m{\"a}l{\"a}inen, Emery~N.~Brown, and~Patrick~L.~Purdon
\end{center}

\subsection{Steady-state covariance of the source process $\vect{\beta}_t$}
\label{subsec:steadystate}

In this section we obtain the steady-state source covariance ${\vect{C}=\mathrm{Cov}(\vect{\beta}_t)}$ of our dynamic source models (Eq. \ref{eq:state}), and show that this covariance matrix is invertible. The matrix ${\vect{F}}$ is stable, i.e., the magnitude of its largest eigenvalue is equal to ${\phi<1}$, since $\vect{F}/\phi$ is a stochastic matrix (it has nonnegative entries with the sum of its rows equal to $1$) and because the largest eigenvalue of a stochastic matrix is $1$. Since ${\vect{F}}$ is stable, the process covariance reaches a steady-state, ${\vect{C} = \lim_{t \rightarrow \infty} \mathrm{Cov}(\vect{\beta}_t)}$, which is given by the solution of the Lyapunov equation ${\vect{C}=\vect{F}\vect{C}\vect{F}' + \vect{Q}}$. Furthermore, since ${\vect{Q}}$ is positive definite and invertible, so it is the source covariance ${\vect{C}}$~\cite{Kailath:2000vg}. Therefore, we can assume that the process ${\vect{\beta}_t}$ has reached a steady-state where its spatial covariance matrix ${\vect{C}=\mathrm{Cov}(\vect{\beta}_t)}$ is time invariant and invertible.

We should note that in our analysis we can parametrized the covariance matrix ${\vect{Q} = [\lambda \mathrm{tr}(\widehat{\vect{\Sigma}}) /n] ^ {-1} \mathrm{diag}(\nu_1,\nu_2,\ldots,\nu_{p})}$, where ${\widehat{\vect{\Sigma}}} = \vect{X}'\vect{X}/n$ is the sample covariance of the rows of $\vect{X}$, ${\lambda > 0}$, and ${\nu_i > 0}$. Doing so would allow us to interpret $\lambda$ as the inverse of the power signal-to-noise ratio ($\mathrm{SNR}^2$). This is because if we assume the matrix $\vect{F}=\phi\vect{I}$, and ${\nu_i = (1-\phi^2)}$ (${i=1, \ldots, p}$), the steady state covariance of $\vect{\beta_t}$ becomes ${\vect{C} = [\lambda\mathrm{tr}(\widehat{\vect{\Sigma}}) /n] ^ {-1}\vect{I}}$. If we define the power signal-to-noise ratio as ${\mathrm{SNR}^2 = \mathrm{E}||\vect{X}\vect{\beta}_t||^2 / \mathrm{E}||\vect{\varepsilon}_t||^2}$, then ${\mathrm{SNR}^2 = \mathrm{tr}(\vect{X}'\vect{X})[\lambda\mathrm{tr}(\widehat{\vect{\Sigma}}) /n] ^ {-1} / n = 1/\lambda}$.

\subsection{The dynamic lead field mapping projects ${[\vect{y}_1', \vect{y}_2', \ldots, \vect{y}_T']'}$ onto ${\vect{\beta}_t}$}
\label{subsec:dynleadisproj}

In this section we show that the observation model for the complete set of measurements ${[\vect{y}_1,\vect{y}_2,\ldots,\vect{y}_T]}$ under the dynamic lead field mapping possesses the desired \textit{signal} plus \textit{noise} structure in which the source vector at time ${t}$ is independent of the \textit{noise} term (Eq.~\ref{eq:dynleadfieldexplicit}). In order to do this, we will show that the matrices ${\vect{P}_{t\pm k,t}}$ that make up the blocks of the dynamic lead field mapping are indeed matrices projecting the corresponding measurement ${\vect{y}_{t\pm k}}$ onto the source vector ${\vect{\beta}_t}$. Doing so would imply that the dynamic lead field mapping achieves the desired independence condition.

We should first point to a few facts about the linear projection ${\vect{P}_{t\pm k,t}\vect{\beta}_t}$ of the measurement vector ${\vect{y}_{t\pm k}}$ onto the source vector ${\vect{\beta}_t}$. By definition, the projection error ${\vect{e}_{t\pm k,t}=\vect{y}_{t\pm k}-\vect{P}_{t\pm k,t}\vect{\beta}_t}$ is uncorrelated with ${\vect{\beta}_t}$\cite{Kailath:2000vg}:

\begin{equation} \label{eq:orthogonalitycond}
    \mathrm{E}\left[(\vect{y}_{t\pm k}-\vect{P}_{t\pm k,t}\vect{\beta}_t)\vect{\beta}_t'\right] = \vect{0}.
\end{equation}

\noindent In addition, since ${\vect{y}_{t\pm k}}$ and ${\vect{\beta}_t}$ are jointly Gaussian, the projection error ${\vect{e}_{t\pm k,t}}$ and ${\vect{\beta}_t}$ are also jointly Gaussian. Therefore, the projection error and the source vector are uncorrelated and jointly Gaussian, and thus are independent. As a result, we can use the definition of the projection error to express any measurement vector ${\vect{y}_{t\pm k}}$ as the sum of the projection ${\vect{P}_{t\pm k,t}\vect{\beta}_t}$ and projection error ${\vect{e}_{t\pm k,t}}$:

\begin{equation*} \label{eq:projection}
    \vect{y}_{t\pm k} = \vect{P}_{t\pm k,t}\vect{\beta}_t + \vect{e}_{t\pm k,t}.
\end{equation*}

\noindent If we do this for all values of ${k}$ we obtain an observation model for the complete time series of measurements where the source vector ${\vect{\beta}_t}$ is indeed independent of the noise term ${[\vect{e}_{1,t}',\vect{e}_{2,t}', \ldots ,\vect{e}_{T,t}']'}$:

\begin{equation*} \label{eq:projectiondynleadfield}
    \begin{bmatrix}
        \vect{y}_1 \\
        \vdots \\
        \vect{y}_{t-1} \\
        \vect{y}_t \\
        \vect{y}_{t+1} \\
        \vdots \\
        \vect{y}_T
    \end{bmatrix}
    =
    \underbrace{
    \begin{bmatrix}
        \vect{P}_{1,t} \\
        \vdots \\
        \vect{P}_{t-1,t} \\
        \vect{P}_{t,t} \\
        \vect{P}_{t+1,t} \\
        \vdots \\
        \vect{P}_{T,t} \\
    \end{bmatrix} \vect{\beta}_t}_{\text{Signal}}
    +
    \underbrace{
    \begin{bmatrix}
        \vect{e}_{1,t} \\
        \vdots \\
        \vect{e}_{t-1,t} \\
        \vect{e}_{t,t} \\
        \vect{e}_{t+1,t} \\
        \vdots \\
        \vect{e}_{T,t}
    \end{bmatrix}}_{\text{Noise}}.
\end{equation*}

\noindent Therefore, if we show that the choices we made for the ${\vect{P}_{t\pm k,t}}$ matrices for the dynamic lead field mapping in Section~\ref{subsec:dynleadfield}, namely,

\begin{align*} 
    \vect{P}_{t,t}    &= \vect{X}, \nonumber \\
    \vect{P}_{t+k,t}  &= \vect{X}\vect{F}^k, \\
    \vect{P}_{t-k,t}  &= \vect{X}\vect{F}_b^k \nonumber,
\end{align*}

\noindent satisfy the orthogonality condition (Eq.~\ref{eq:orthogonalitycond}), it would follow that the \textit{signal} and \textit{noise} portions in Equation~\eqref{eq:dynleadfieldexplicit} are independent.

In the case where the projection matrix corresponds to the present observation ${(k=0)}$, we substitute Equation~\eqref{eq:obsmodel} in Equation~\eqref{eq:orthogonalitycond}, and note that setting ${\vect{P}_{t,t}=\vect{X}}$ achieves the desired orthogonality condition since the source vector ${\vect{\beta}_t}$ is independent of the measurement noise ${\vect{\varepsilon}_t}$:

\begin{align*} 
    \mathrm{E}\left[(\vect{y}_t-\vect{P}_{t,t}\vect{\beta}_t)\vect{\beta}_t'\right]
        &= \mathrm{E}\left[(\vect{X}\vect{\beta}_t + \vect{\varepsilon}_t-\vect{P}_{t,t}\vect{\beta}_t)\vect{\beta}_t'\right] \nonumber \\
        &= \mathrm{E}\left[\vect{\varepsilon}_t\vect{\beta}_t'\right] \nonumber \\
        &= \vect{0}.
\end{align*}

To analyze the case where the measurement corresponds to a future observation, we should first make note of two facts in relation to the source vector process. The first is that, by definition, the input process to the source dynamics is independent across time, i.e., ${\vect{\omega}_t}$ is independent of ${\vect{\omega}_{t \pm j}}$ for ${j \in [1, 2, \ldots]}$. The second is that given the recursive definition of the source process (Eq.~\ref{eq:state}), the source vector at a particular time ${t}$ is a function of only the present input ${\vect{\omega}_t}$ and past inputs ${\vect{\omega}_{t-j},\: j\in[1,2,\ldots]}$. Because of these two facts, the source vector a time ${t}$, ${\vect{\beta}_t}$, is independent of the future inputs ${\vect{\omega}_{t+j},\:j\in[1,2,\ldots]}$.

With this result at hand, we proceed to deduce the future projection matrices ${\vect{P}_{t+k,t}, \:k\in[1,2,\ldots]}$ by using the forward iteration in Equation~\eqref{eq:kfuturepred}: ${\vect{\beta}_{t+k} = \vect{F}^k \vect{\beta}_t + \sum_{j=1}^k \vect{F}^{k-j}\vect{\omega}_{t+j}}$. We substitute Equations~\eqref{eq:obsmodel} in Equation~\eqref{eq:orthogonalitycond}, then substitute Equation~\eqref{eq:kfuturepred} in the obtained result, and note that by setting ${\vect{P}_{t+k,t}=\vect{X}\vect{F}^k}$ we achieve the orthogonality condition since ${\vect{\beta}_t}$ is independent of both ${\vect{\varepsilon}_{t+k}}$, and ${\vect{\omega}_{t+j},\:j\in[1,2,\ldots]}$:

\begin{align*} 
    \mathrm{E}[(\vect{y}_{t+k}-\vect{P}_{t+k,t}\vect{\beta}_t)\vect{\beta}_t']
        &= \mathrm{E}[(\vect{X}\vect{\beta}_{t+k} + \vect{\varepsilon}_{t+k}-\vect{P}_{t+k,t}\vect{\beta}_t)\vect{\beta}_t'] \nonumber \\
        &= \mathrm{E}[(\vect{X}\vect{F}^k\vect{\beta}_t + {\textstyle\sum\limits_{j=1}^k} \vect{X}\vect{F}^{k-j}\vect{\omega}_{t+j} + \vect{\varepsilon}_{t+k}-\vect{P}_{t+k,t}\vect{\beta}_t)\vect{\beta}_t'] \nonumber \\
        &= \mathrm{E}[({\textstyle\sum\limits_{j=1}^k} \vect{X}\vect{F}^{k-j}\vect{\omega}_{t+j} + \vect{\varepsilon}_{t+k})\vect{\beta}_t'] \nonumber \\
        &= \vect{0}.
\end{align*}

To analyze the past observations ${\vect{y}_{t-k}}$, ${k \in [1,2,\ldots]}$, we proceed similarly to the case of the future projections but use instead the \textit{backwards Markovian} representation (Eq.~\ref{eq:backwardsmarkov}) and its ${k}$-step past iteration (Eq.~\ref{eq:kpastpred})\footnote{We should point out that directly reversing the time direction of the equation defining the source spatiotemporal dynamics (Eq.~\ref{eq:state}), i.e., ${\vect{\beta}_{t-1} = \vect{F}^{-1}\vect{\beta}_t - \vect{F}^{-1}\vect{\omega}_t}$, instead of using the \textit{backwards Markovian} model does not yield the desired orthogonality condition. This is because the ${k}$-step backwards iteration resulting from this direct time reversal, namely, ${\vect{\beta}_{t-k} = \vect{F}^{-k}\vect{\beta}_t - \sum_{j=0}^{k-1} \vect{F}^{j-k}\vect{\omega}_{t-j}}$, contains the terms ${\vect{\omega}_t}$ and ${\vect{\omega}_{t-j}}$ which are not independent of the source vector ${\vect{\beta}_t}$.}: ${\vect{\beta}_{t-k} = \vect{F}_b^k \vect{\beta}_t + \sum_{j=1}^k \vect{F}_b^{k-j}\vect{\omega}_{t-j}^b}$. We should note that in the \textit{backwards Markovian} representation (Eq.~\ref{eq:backwardsmarkov}), the source vector ${\vect{\beta}_t}$ is a function of only the present ${\vect{\omega}_t^b}$ and future inputs ${\vect{\omega}_{t+j}^b}$, ${j\in[1,2,\ldots]}$. In addition, given that the backwards input process ${\vect{\omega}_t^b}$ is independent across time, in this equivalent representation the source vector ${\vect{\beta}_t}$ is independent of the past inputs ${\vect{\omega}_{t-j}^b}$, ${j\in[1,2,\ldots]}$. Using this result, we can readily obtain the past projection matrices. We use Equations~\eqref{eq:obsmodel},~\eqref{eq:orthogonalitycond}, and~\eqref{eq:kpastpred} to deduce that by setting ${\vect{P}_{t-k,t}=\vect{X}\vect{F}_b^k}$ we obtain the desired orthogonality condition since ${\vect{\beta}_t}$ is independent of both ${\vect{\varepsilon}_{t-k}}$, and ${\vect{\omega}_{t-j}^b,\:j\in[1,2,\ldots]}$:

\begin{align*} 
    \mathrm{E}[(\vect{y}_{t-k}-\vect{P}_{t-k,t}\vect{\beta}_t)\vect{\beta}_t']
        &= \mathrm{E}[(\vect{X}\vect{\beta}_{t-k} + \vect{\varepsilon}_{t-k}-\vect{P}_{t-k,t}\vect{\beta}_t)\vect{\beta}_t'] \nonumber \\
        &= \mathrm{E}[(\vect{X}\vect{F}_b^k\vect{\beta}_t + {\textstyle\sum\limits_{j=1}^k} \vect{X}\vect{F}_b^{k-j}\vect{\omega}_{t-j}^b + \vect{\varepsilon}_{t-k}-\vect{P}_{t-k,t}\vect{\beta}_t)\vect{\beta}_t'] \nonumber \\
        &= \mathrm{E}[({\textstyle\sum\limits_{j=1}^k} \vect{X}\vect{F}_b^{k-j}\vect{\omega}_{t-j}^b + \vect{\varepsilon}_{t-k})\vect{\beta}_t'] \nonumber \\
        &= \vect{0}.
\end{align*}

\subsection{Ethics statement}
\label{subsec:ethics}

Human studies were approved by the Human Research Committee of the Massachusetts General Hospital, Boston, MA.

\subsection{Data description and preprocessing}
\label{subsec:datapreprocessing}

In our analysis, the lead field matrix ${\vect{X}}$ was computed with the \textit{MNE} software using an MRI-based boundary element model from a human subject~\cite{Hamalainen:1993ws,Dale:1993wo}, with ${p=5124}$ dipole sources oriented normal to the cortical surface. This arrangement yielded an average distance between nearest neighbors of ${\sim 6.2 \:\mathrm{mm}}$, resulting in a model that is consistent with intracranial electrophysiology and neuroanatomy~\cite{Bullock:1995wy,Destexhe:1999tc,Leopold:2003va,Nunez:1995vc}. Due to the difference in orders of magnitude and physical units in the elements of the lead field matrix ${\vect{X}}$ corresponding to planar gradiometers and magnetometers, we restricted our analysis to the 204 planar gradiometers of the Neuromag Vectorview system at the Massachusetts General Hospital. The $\vect{F}$ matrix was specified using the nearest-neighbor dynamic formulation described by Equations~\eqref{eq:sourcedyn} and~\eqref{eq:state}, with the stability parameter ${\phi = 0.95}$. The state input covariance matrix ${\vect{Q}}$ was estimated from \textit{mu}-rhythm data from one subject using the \textit{dMAP-EM} algorithm presented in~\cite{Lamus:2012gp}. The \textit{mu}-rhythm originates from motor and somatosensory cortices, and consists of synchronous oscillations with 10 and 20-Hz components. Data were collected from one subject using a 306-channel Neuromag Vectorview MEG system at the Massachusetts General Hospital. The signals were acquired at 601 Hz with a bandwidth of 0.1 to 200 Hz, and downsampled to 204.8 Hz for subsequent analysis. In the \textit{STS} model, the elements for the temporal covariance matrix $\vect{\Gamma}$ were set as suggested in~\cite{Friston:2008jr}: ${\gamma_{t \pm k,t} = \sum_{j=1}^T \exp \{ -1/2 [(t \pm k-j)^2 + (j-t)^2] \Delta^{-2} \psi^{-2} \}}$, where $\Delta = 4\times10^{-3} \: \text{s}$ and $\psi = 204.8\: \text{Hz}$.

\end{document}